\documentclass[aps,prl,twocolumn,showpacs,preprintnumbers,floatfix,amsmath,superscriptaddress]{revtex4-1}
\usepackage{graphicx}
\usepackage{dcolumn}
\usepackage{subfigure}
\usepackage{wrapfig}
\usepackage{cancel}
\usepackage{color}
\usepackage{bm}
\usepackage{verbatim}
\usepackage{comment}
\usepackage{epstopdf}
\usepackage{gensymb}
\usepackage{float}
\usepackage{textcomp}
\usepackage{mathtools}
\usepackage{MnSymbol}

\begin{document}

\title{Fidelity Study of Superconductivity in Extended Hubbard Models}
\author{N. Plonka}
\email{Author to whom correspondence should be addressed: nachumplonka@gmail.com}
\affiliation{Department of Applied Physics, Stanford University, Stanford, California 94305, USA}
\affiliation{Stanford Institute for Materials and Energy Sciences, SLAC National Accelerator Laboratory, Menlo Park, California 94025, USA}
\author{C.J. Jia}
\affiliation{Department of Applied Physics, Stanford University, Stanford, California 94305, USA}
\affiliation{Stanford Institute for Materials and Energy Sciences, SLAC National Accelerator Laboratory, Menlo Park, California 94025, USA}
\author{Y. Wang}
\affiliation{Department of Applied Physics, Stanford University, Stanford, California 94305, USA}
\affiliation{Stanford Institute for Materials and Energy Sciences, SLAC National Accelerator Laboratory, Menlo Park, California 94025, USA}
\author{B. Moritz}
\affiliation{Stanford Institute for Materials and Energy Sciences, SLAC National Accelerator Laboratory, Menlo Park, California 94025, USA}
\affiliation{Department of Physics and Astrophysics, University of North Dakota, Grand Forks, ND 58202, USA}
\author{T. P. Devereaux}
\email{Author to whom correspondence should be addressed: tpd@slac.stanford.edu}
\affiliation{Stanford Institute for Materials and Energy Sciences, SLAC National Accelerator Laboratory, Menlo Park, California 94025, USA}
\affiliation{Geballe Laboratory for Advanced Materials, Stanford University, Stanford, California 94305, USA}

\begin{abstract}

The Hubbard model with local on-site repulsion is generally thought to possess a superconducting ground-state for appropriate parameters,  but the effects of more realistic long-range Coulomb interactions have not been studied extensively.  We study the influence of these interactions on superconductivity by including nearest and next-nearest neighbor extended Hubbard interactions in addition to the usual on-site terms.  Utilizing numerical exact diagonalization, we analyze the signatures of superconductivity in the ground states through the fidelity metric of quantum information theory.  We find that nearest and next-nearest neighbor interactions have thresholds above which they destabilize superconductivity regardless of whether they are attractive or repulsive, seemingly due to competing charge fluctuations.

\end{abstract}


\maketitle

The role of strong electronic correlations on unconventional superconductivity (SC) remains an important open question \cite{Scalapino2012}.  The two-dimensional single-band Hubbard model with an on-site repulsive Coulomb interaction has been well-studied as the simplest effective model which can capture the influence of correlations on the electronic properties.  It is generally accepted that it possesses a superconducting ground state with $d$-wave symmetry upon carrier doping (changing the chemical potential); however, only a few studies can provide an exact solution, including a thermodynamic limit calculation limited to asymptotically small coupling \cite{Kohn1965,Raghu2010}.  In the intermediate to strong coupling regime, studies include numerically exact simulations \cite{Maier2005,Maier2005a,Haule2007}, $t-J$ model calculations \cite{Anderson1987,Lee1988}, and exact diagonalization on finite clusters \cite{Rigol2009, Rigol2009b, Jia2011}. 

While the on-site Coulomb terms in the doped Hubbard model are sufficient for producing a superconducting ground state, one would like to understand the effects of more realistic, long-range Coulomb interactions.  Often, Hubbard models are used to approximate real materials such as cuprates, so parameter choices are typically  renormalizations of their bare values.  As such, it makes sense to consider both repulsive and attractive extended interactions of various magnitudes \cite{Feiner1995}.  Naively, repulsive (attractive) nearest neighbor interactions are expected to suppress (enhance) SC, since $d$-wave pairs comprise spins on nearest neighbor sites.  Conversely, attractive \emph{next}-nearest neighbor interactions favor diagonally adjacent spins but seemingly not SC's nearest neighbor configuration, whereas repulsive interaction may enhance SC by suppressing diagonal adjacency.  Calculations of repulsive nearest-neighbor interactions find that it indeed suppresses SC, but this does not mean that SC is destabilized.  Although some contend that arbitrarily small interactions will destabilize SC \cite{Alexandrov2011}, several calculations show that SC remains stable until interactions reach some threshold \cite{Gazza1999,Onari2004,Raghu2012,Plakida2013,Senechal2013}. On the other hand, attractive nearest-neighbor interactions are found in Hartree-Fock calculations to enhance SC above some threshold when considered alone \cite{Micnas1988}.  However, one should also consider spin and charge correlations driven by these interactions, as they may enhance \cite{Martin2001} or compete \cite{Su2001} with SC.

Here, we apply the fidelity metric from quantum information theory to explore the influence of extended Hubbard interactions on SC.  This method has been used to investigate classical as well as quantum critical behavior in various systems \cite{Zanardi2006,Chen2007,CamposVenuti2007,You2007,Buonsante2007,Cozzini2007,CamposVenuti2008,Abasto2008,Gu2008,Garnerone2009,Garnerone2009a} and is a clear test of SC \cite{Rigol2009,Rigol2009b,Jia2011}.   Utilizing numerical exact diagonalization, we find that nearest and next-nearest neighbor interactions have thresholds above which they destabilize SC regardless of whether they are attractive or repulsive.

\begin{figure}
\begin{center}
\includegraphics[width=\columnwidth, viewport=0in 1in 13in 12in]{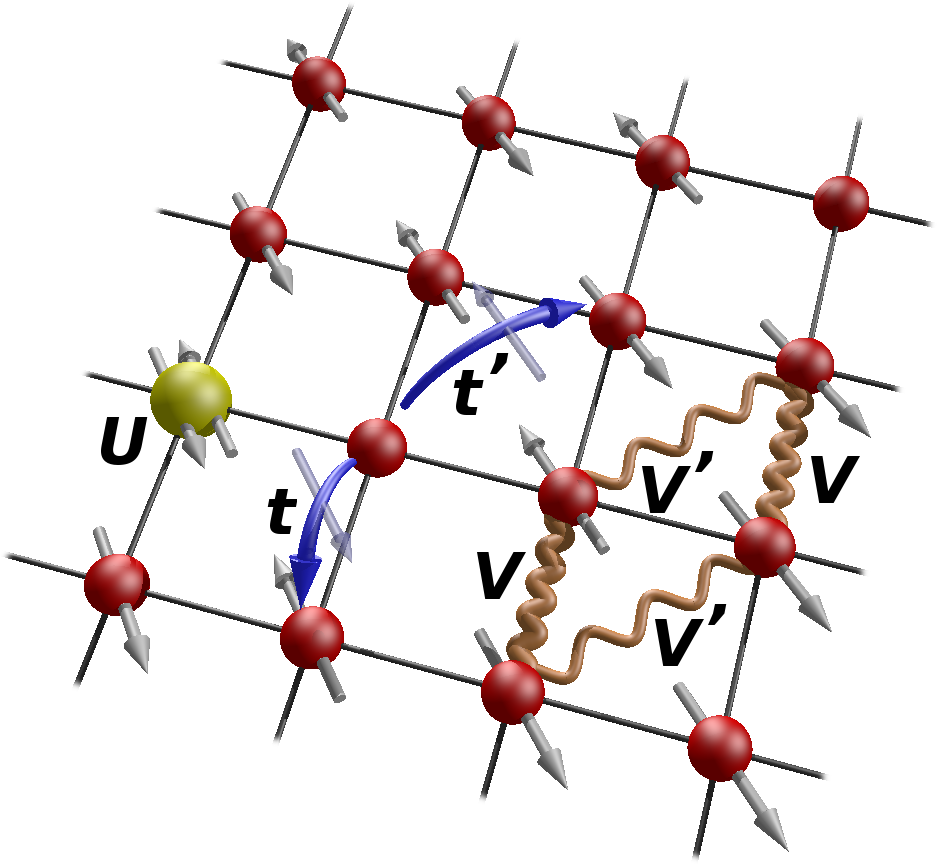}   
\end{center}
\caption{(Color online) The operations of the kinetic and interaction Hamiltonian terms on spins: $t$ and $t'$ are nearest and next-nearest neighbor hopping, resp, $U$ is the onsite interaction, and $V$ and $V'$ are next-nearest neighbor interactions.  The lattice is a 16B Betts cluster with periodic boundary conditions, with atoms at the intersections.}
\label{fig:Hamiltonian}
\end{figure}

\emph{Theory. - } The two-dimensional extended Hubbard model in real space comprises kinetic and interaction terms, depicted schematically in Fig. \ref{fig:Hamiltonian}.  The kinetic portion of the Hamiltonian is
\begin{equation}
H_\mathrm{kin}=-t \sum_{\langle ij \rangle \sigma} (c_{i\sigma}^\dagger c_{j\sigma} + c_{j\sigma}^\dagger c_{i\sigma}) + t' \sum_{\llangle il \rrangle \sigma} (c_{i\sigma}^\dagger c_{l\sigma} + c_{l\sigma}^\dagger c_{i\sigma}) ,
\label{Eq:H0}
\end{equation}  
where $c_{i\sigma}^\dagger$ ($c_{i\sigma}$) creates (annihilates) a fermion at site $i$ with spin $\sigma$. $\langle \rangle$ and $\llangle \rrangle$ denote nearest and next-nearest neighbor sites, resp.  Throughout, we set $t=1$. The interacting portion of the Hamiltonian includes extended interactions up to next-nearest neighbors:
\begin{align}
H_\mathrm{int} = &U \sum_{i} c_{i\uparrow}^\dagger c_{i\downarrow}^\dagger c_{i\downarrow} c_{i\uparrow} + V \sum_{\langle ij \rangle \sigma\sigma'} c_{i\sigma}^\dagger c_{j\sigma'}^\dagger c_{j\sigma'} c_{i\sigma} \nonumber \\
 &+ V' \sum_{\llangle il \rrangle \sigma\sigma'} c_{i\sigma}^\dagger c_{l\sigma'}^\dagger c_{l\sigma'} c_{i\sigma}.
\label{Eq:Hint}
\end{align}  

We determine the ground state wavefunction for the model using exact diagonalization.  The Hamiltonian matrix was diagonalized using the Parallel Arnoldi PACKage (PARPACK) \cite{Lehoucq1998}, which provides an efficient method to obtain a set of properly orthonormalized eigenvectors and is more numerically stable compared with standard Lanczos techniques.  Due to the exponential growth of the Hilbert space with cluster size, we work with a conserved electron number (canonical ensemble), fix the total momentum (translational symmetry), and work in the total spin 0 sector.  In a past fidelity study of the 16B Betts cluster \cite{Jia2011}, the Hubbard model with $U \sim W$ was found to have a superconducting ground state when doped with 4 holes (25$\%$ filling)  and with periodic boundary conditions (Fig. \ref{fig:Hamiltonian}).  Therefore, we study this cluster for the effects of extended interactions on SC.

Having determined the ground state, one can understand the physical properties of the model using correlation functions.  In our study, we are interested in the superconducting correlator, and the spin and charge correlations are also basic for our understanding.  For $d$-wave SC, the pair density matrix \cite{Gorkov1958,Yang1962,Rigol2009,Jia2011} between sites $i,j$ at any separation distance in the ground state $| \Phi_0 \rangle$ is $P_{ij}^\mathrm{SC} = \langle \Phi_0 | D_i^\dagger D_j | \Phi_0 \rangle$, where $D_i = \Delta_{i,i+\hat{x}} - \Delta_{i,i+\hat{y}}$, $\Delta_{ij}=c_{i\uparrow}c_{j\downarrow} + c_{j\uparrow}c_{i\downarrow}$.  The charge density matrix is $P_{ij}^\mathrm{charge} =\langle \Phi_0 | \rho_i \rho_j | \Phi_0 \rangle - \langle \Phi_0 | \rho_i | \Phi_0 \rangle ^2$, where $\rho_i=c_{i\uparrow}^\dagger c_{i\uparrow} + c_{i\downarrow}^\dagger c_{i\downarrow}$.  The spin density matrix is $P_{ij}^\mathrm{spin} =\langle \Phi_0 | S_i S_j | \Phi_0 \rangle$, where $S_i=c_{i\uparrow}^\dagger c_{i\uparrow} - c_{i\downarrow}^\dagger c_{i\downarrow}$. 

Specifically, these correlators are used to construct charge and spin structure factors, which we analyze as a function of wave vector.
\begin{align}
N(\mathbf{q})&=(1/N) \sum_{j} P_{0j}^\mathrm{charge} e^{i \mathbf{q} \cdot \mathbf{r}_j} \label{Eq:N} \\ 
S(\mathbf{q})&=(1/N) \sum_{j} P_{0j}^\mathrm{spin} e^{i \mathbf{q} \cdot \mathbf{r}_j}
\end{align}
where $N$ is the number of sites, $\mathbf{q}$ is the wave vector, and $\mathbf{r}_j$ is the position of site $j$.  A structure factor to measure relative SC is similarly constructed with $P_{ij}^\mathrm{SC}$, but it is preferable to use an absolute measure of the presence of SC.

To investigate SC in Hubbard models, we calculate the fidelity, which was first used widely in quantum information to provide the criterion for distinguishing between quantum states \cite{Steane1998,Bennett2000}.  This quantity has now been utilized increasingly in general physics for investigating classical as well as quantum critical behavior in various systems \cite{Zanardi2006,Chen2007,CamposVenuti2007,You2007,Buonsante2007,Cozzini2007,CamposVenuti2008,Abasto2008,Gu2008,Garnerone2009,Garnerone2009a}.  This metric measures the ground state wavefunction's rate of change and shows whether the ground state adiabatically connects to a superconducting state.  Recently, it was used to demonstrate SC in the $t$-$J$ \cite{Rigol2009,Rigol2009b} and on-site Hubbard models \cite{Jia2011}.   

This technique involves adding an additional term to the Hamiltonian:
\begin{equation}
H_\mathrm{SC} = - \frac{\lambda_\mathrm{SC}}{N} \sum_{ij} D_i^\dagger D_j; \label{Eq:SC}
\end{equation}
The full Hamiltonian has the form $H(\lambda_\mathrm{SC}) = H_\mathrm{kin} + H_\mathrm{int} + H_\mathrm{SC}$.  The fidelity metric is defined as
\begin{equation}
g(\lambda_\mathrm{SC},\delta\lambda) = (2/N)(1-F(\lambda_\mathrm{SC},\delta\lambda))/\delta\lambda^2,
\end{equation}
where the fidelity $F=\langle \Phi_0(\lambda_\mathrm{SC}) | \Phi_0(\lambda_\mathrm{SC}+\delta\lambda) \rangle$.  For finite-size systems as in this work, a change in the ground state's symmetry occurs via a quantum critical crossover (QCC) that appears as a broad peak in $g$.  This provides a signal of strong SC: The ground state of $H(\lambda_\mathrm{SC})$ at large $\lambda_\mathrm{SC}$ is expected to possess long-range SC with a correlation length at least of order the cluster size.  If $g$ displays no QCC as one reduces $\lambda_\mathrm{SC}$, then this SC persists in the Hamiltonian of interest down to $\lambda_\mathrm{SC}=0$.  

To substantiate the results of a fidelity analysis, we evaluate the ratio between the largest and next largest eigenvalues of the density matrices $P_{0j}^\mathrm{SC}$ and $P_{0j}^\mathrm{charge}$, referred to as $R$ ratios \cite{Penrose1956,Yang1962,Rigol2009,Rigol2009b,Jia2011}.   These eigenvalues are proportional to the structure factors for all values of $\mathbf{q}$ (Eq. \ref{Eq:N} and following text).  Therefore, for charge correlations, if $R_\mathrm{charge}>1$, there is a $\mathbf{q}$ for which the correlations are strongest, and $R_\mathrm{charge}$ is the factor by which they are larger than the next strongest correlations.  If $R_\mathrm{charge}=1$, then there is no dominant $\mathbf{q}$.  For SC, only the familiar $\mathbf{q}=0$ correlations may be present in our model, so this is the dominant correlation for cases where $R_\mathrm{SC}>1$.  Like correlators, the magnitude of $R>1$ is not meaningful by itself, but its trends as a function of interaction parameters provide confirmation for the fidelity results.

\begin{figure}
\begin{center}
\includegraphics[width=\columnwidth, viewport=0in 0.5in 8.25in 8.5in]{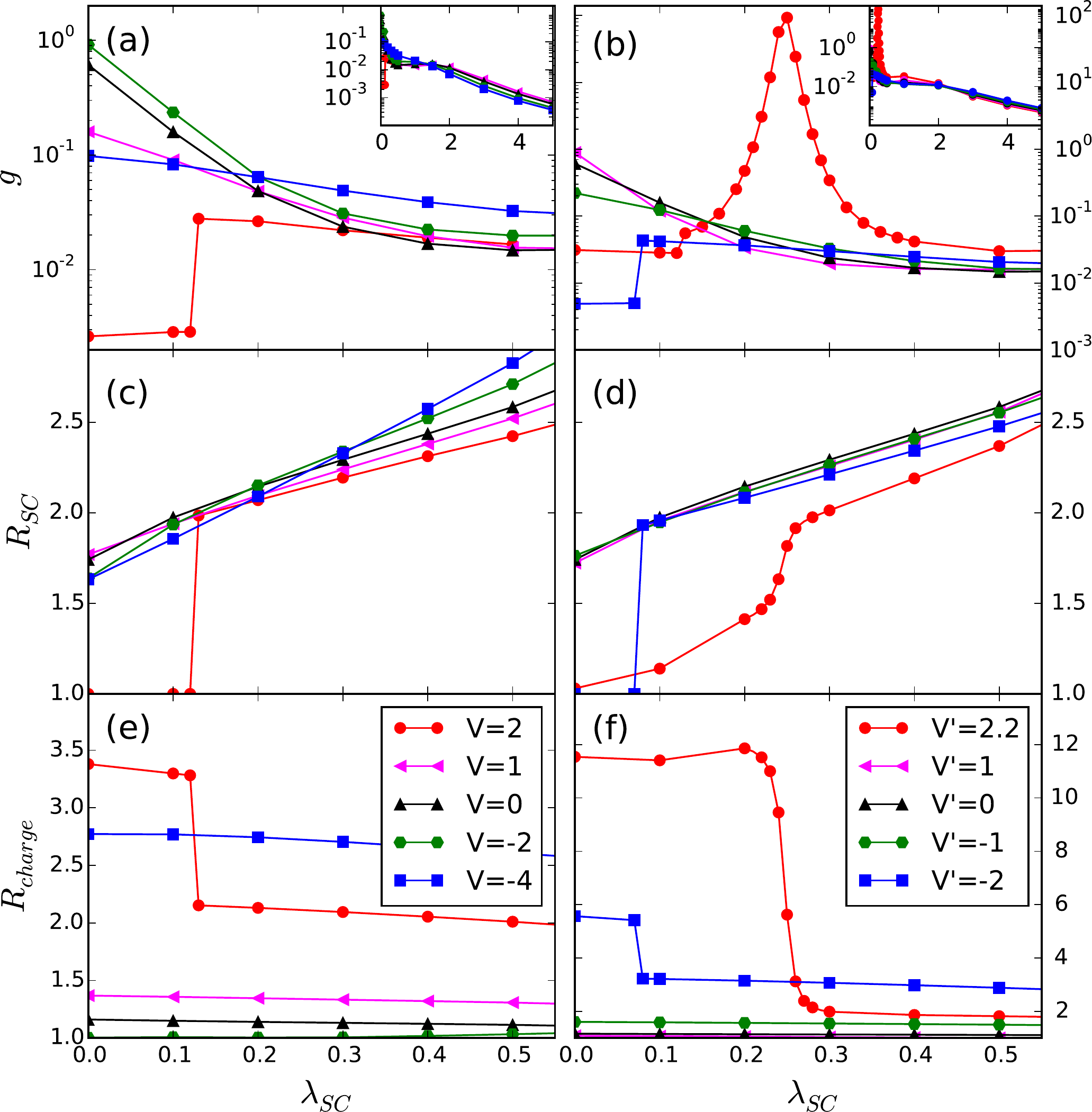}  
\end{center}
\caption{(Color online) The semi-logarithmic plot of fidelity metric vs. $\lambda_{\mathrm{SC}}$ for repulsive and attractive (a) nearest-neighbor and (b) next-nearest neighbor interactions on superconductivity with 4 doped holes.  The insets show the fidelity metric over a larger range of $\lambda_{\mathrm{SC}}$ to confirm that it is smooth without peaks for $\lambda_{\mathrm{SC}}>0.5$.  Also shown are $R_\mathrm{SC}$ ratios for (c) nearest-neighbor and (d) next-nearest neighbor interactions and $R_\mathrm{charge}$ density matrix eigenvalue ratios for the same (e,f).}
\label{fig:fidelity}
\end{figure}

\emph{Results. - } We show the effects of nearest and next-nearest neighbor interactions - both repulsive and attractive - on superconductivity.  The SC is clearest for the parameters $U=8, t'=-0.3$ \cite{Jia2011}, so we employ the same set here to analyze the effects of extended interactions on SC.  To stabilize the numerics, steps of $\delta\lambda=10^{-4}$ prove to be sufficient.

Figure \ref{fig:fidelity}(a) shows the fidelity metric for nearest neighbor interactions.  For large $\lambda_{\mathrm{SC}}$, the state is expected to to possess long-range SC with a correlation length at least of order the cluster size.  Starting with a strong pairing field, $\lambda_{\mathrm{SC}}=5$, and decreasing $\lambda_{\mathrm{SC}}$, the fidelity metric $g$ increases monotonically (see inset of Fig. \ref{fig:fidelity}(a)).  This indicates an increased rate of change in the ground state as it approaches a potential QCC.  For $V=0$, $g$ continues to increase monotonically and does not peak down to $\lambda_{\mathrm{SC}}=0$; The ground state is still superconducting at $\lambda_{\mathrm{SC}} = 0$, as found in the previous study \cite{Jia2011}.  

Adding nearest neighbor interactions, we find that attractive and repulsive interactions have thresholds above which they suppress SC.  The range of interactions shown is chosen to reach or cross these thresholds.  For the repulsive case, $V=1$ is similar to $V=0$ but for $V=2$, the QCC peak, which defines $\lambda_{\mathrm{SC}}^*$, is at 0.12; at $\lambda_{\mathrm{SC}} = 0$ the system is no longer superconducting.  Note that although this peak includes a sharp jump, the ground state's total momentum remains at zero for all $\lambda_{\mathrm{SC}}$ shown.  For attractive $V<0$, $g$ increases with decreasing $\lambda_{\mathrm{SC}}$ even for interactions as large as $V=-4$ but with a much smaller slope than for smaller attractive interaction.  For larger interactions $\lambda_{\mathrm{SC}}^*>0$ as SC is destabilized (not shown).  This contradicts the intuition that attractive and repulsive interactions have opposite effects on SC.

Similarly for next-nearest neighbor interactions in Fig. \ref{fig:fidelity}(b), we find that attractive and repulsive interactions have destabilization thresholds.  For $|V'|=1$, $g$ increases with decreasing $\lambda_{\mathrm{SC}}$ (see inset of Fig. \ref{fig:fidelity}(b)), indicating that the system remains superconducting.  However, for larger $V'=-2, 2.2$, one observes $\lambda_{\mathrm{SC}}^*>0$; the system is no longer superconducting, although the total momentum is unchanged.  (At $V'=2$ the peak is still at $\lambda_{\mathrm{SC}}^* \leq 0$.)  Contrary to intuition, nearest and next-nearest neighbor interactions all destabilize SC beyond some threshold, regardless of whether they are attractive or repulsive.  All of these destabilizing cases with $\lambda_{\mathrm{SC}}^*>0$ are confirmed by level crossings in the ground state energy at $\lambda_{\mathrm{SC}}^*$ (not shown).  It is possible that there are competing phenomena stabilized by these interactions.

The fidelity results for nearest and next-nearest neighbor interactions are further confirmed by $R_\mathrm{SC}$ ratios in Figs. \ref{fig:fidelity}(c,d), resp.  In the cases where $V$ or $V'$ destabilize SC and the fidelity metric peaks at $\lambda_{\mathrm{SC}}^*>0$, $R_\mathrm{SC}$ drops sharply as $\lambda_{\mathrm{SC}}$ crosses below $\lambda_{\mathrm{SC}}^*$, corroborating the suppression of SC.  Additionally, the observation that $R_\mathrm{SC}=1$ at $\lambda_{\mathrm{SC}} = 0$ clearly indicates a non-superconducting system. Conversely, for the $V, V'$ cases with stable SC,  $R_\mathrm{SC}$ evolves smoothly for $\lambda_{\mathrm{SC}}>0$ just like the fidelity metric, implying no destabilization of SC in this range.

To explore the possibility of phenomena competing with SC, we consider charge fluctuations (CF), which are analyzed via the $R_\mathrm{charge}$ ratios for nearest and next-neighbor interactions in Figs. \ref{fig:fidelity}(e,f), resp.  In all cases where SC is destabilized, as $R_\mathrm{SC}$ drops sharply at $\lambda_{\mathrm{SC}}^*$, $R_\mathrm{charge}$ increases sharply. This suggests that the increasing CF destabilizes SC.  Consistently, in cases of stable SC, both $R_\mathrm{SC}$ and $R_\mathrm{charge}$ evolve smoothly.

\begin{figure}
\begin{center}
\includegraphics[width=\columnwidth, viewport=0in 0.5in 8.5in 7in]{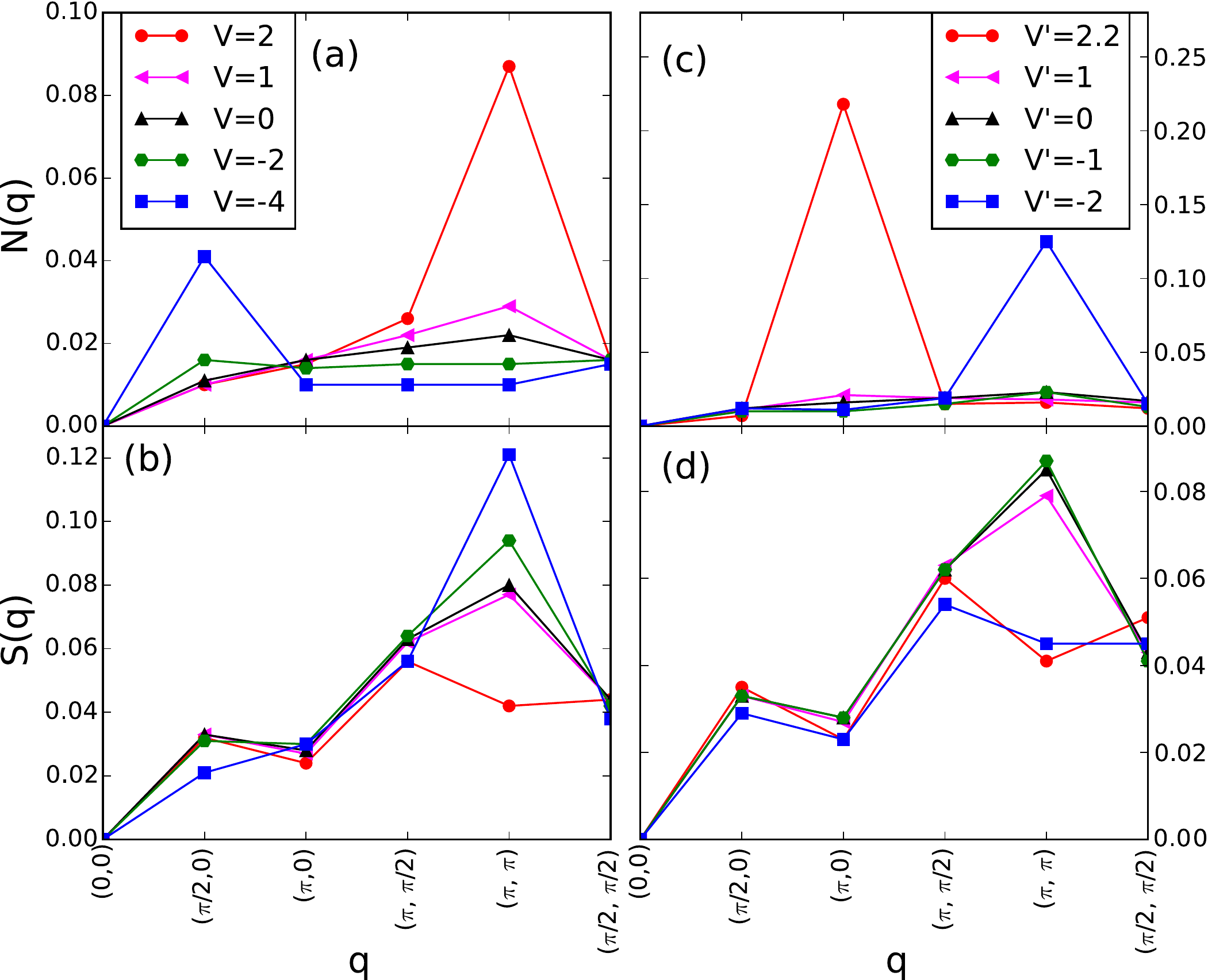} 
\end{center}
\caption{(Color online) The (a) charge and (b) spin structure factors for all wave vectors $\mathbf{q}$ in the cluster at which they can be distinct in our model, for nearest neighbor interactions, and (c,d) the same for next-nearest neighbor interactions.}
\label{fig:NS}
\end{figure}

To find the wave vector of the CF for nearest neighbor interactions, Fig. \ref{fig:NS}(a) displays the charge structure factor $N(\mathbf{q})$.  At $V=2$, where SC is destabilized, $N(\mathbf{q})$ clearly peaks for CF with wave vector $\mathbf{q}=(\pi,\pi)$.  For the other values of $V$ that do not destabilize SC, $(\pi,\pi)$ CF is suppressed.  $V<0$ favors $(\frac{\pi}{2},0)$ CF, which coexists with SC until it is destabilized for $V<-4$ (not shown).  Both attractive and repulsive interactions favor CF that seems to destabilize SC, but they have different wave vectors.

We also consider the interplay between the spin structure factor $S(\mathbf{q})$ in Fig. \ref{fig:NS}(b) and superconductivity.  At $V=0$, the presence of SC is accompanied by $(\pi,\pi)$ spin fluctuations (SF).  These are suppressed along with SC at $V=2$ in favor of CF, whereas they increase for $V<0$ where SC survives.  It is CF, not SF, that appear to compete with SC at larger values of $V$.

For next-nearest neighbor interactions, the CF is found via the charge structure factor in Fig. \ref{fig:NS}(c).  At $V'=-2$, where SC is destabilized, the structure factor $N(\mathbf{q})$ peaks for CF with wave vector $\mathbf{q}=(\pi,\pi)$, while for $V'=2.2$ it peaks at $\mathbf{q}=(\pi,0)$.  For the other values of $V'$ that do not destabilize SC, no CF dominates.  Both the attractive and repulsive interactions favor CF that destabilize SC, but with different wave vectors.  Conversely, in Fig. \ref{fig:NS}(d), the spin structure factor shows that $\mathbf{q}=(\pi,\pi)$ SF follow the behavior of SC and are thus suppressed at larger $|V'|$.  

\emph{Discussion and Conclusion. - } Utilizing numerical exact diagonalization, we analyzed the signatures of superconductivity through the fidelity metric, $R$ ratios and structure factors.  Consistent with intuition and past studies \cite{Gazza1999,Onari2004,Raghu2012,Plakida2013,Senechal2013}, repulsive nearest neighbor interactions destabilize SC beyond some threshold seemingly due to competing CF.  However, we find that this trend holds true for all nearest and next-nearest neighbor interactions, whether repulsive or attractive.

These trends can be understood by simple considerations.  The nearest neighbor repulsion and next-nearest neighbor attraction do not favor neighboring charge (see Fig. \ref{fig:Hamiltonian}) and therefore drive $(\pi,\pi)$ CF, consistent with lower dimensional \cite{Hirsch1984,Raghu2012} and weak coupling studies \cite{Zhang1989,Onari2004,Davoudi2007}.  The diagonally adjacent charge favored by $(\pi,\pi)$ CF is incompatible with the nearest-neighbor spins of $d$-wave SC.  The next-nearest neighbor repulsion favors $(\pi,0)$ CF for larger interactions.  Even this wave vector may be incompatible with SC because it is somewhat restrictive on neighboring spins.  Even if the obtained wave vectors may be subject to finite size limitations of the cluster, they are consistent with their driving interactions and other studies \cite{Wohlfeld2007,Wohlfeld2010,Wang2014,Allais2014}.

Even attractive nearest neighbor interactions, which may be expected to favor $d$-wave SC, will destabilize it for large enough interactions due to CF.  This can be understood by considering the strong coupling limit, $t/U, t/V \rightarrow 0$.  For attractive $|V| \gg U$, the electrons will gather as closely as possible.  In our 16-site model with twelve electrons, the optimal configurations have six doubly-occupied sites with at most seven nearest neighbor bonds, by occupying a $2 \times 3$ block for example.  The structure factors of these configurations peak at $N(\frac{\pi}{2},0)$ (or equally $N(0,\frac{\pi}{2})$) because their periodicity equals the size of the lattice.  Their energy is $E_{|V| \gg U} = 28V + 6U$.  Conversely, in the $|V| \ll U$ limit, the sites are rarely doubly-occupied.  Twelve out of sixteen singly-occupied sites have between sixteen and twenty occupied bonds, resulting in an energy $E_{|V| \ll U}$ between $16V$ and $20V$.  Setting $E_{|V| \gg U} = E_{|V| \ll U}$, these energies cross at $V$ between $-U/2$ and $-3U/4$, consistent with $V=-U/2$ at which we obtained the destabilization of SC by CF.  Furthermore, if such a strong coupling picture is appropriate in the thermodynamic limit, one may expect some threshold $|V| \lesssim U$ above which charge gathers together and SC is destabilized.

In this work, we are able to access a broad phase space of extended interactions in a well-controlled manner via numerical simulations.  The results are suggestive for the rich phase diagrams of strongly correlated materials, which may have varying effective extended interactions and screening lengths.  These can give rise to or influence charge, spin, and superconducting behavior.

\begin{acknowledgments}
We would like to acknowledge helpful discussions with M. Claassen, S. Kivelson, Y. Kung, R. Laughlin, A. Maharaj, S. Raghu, S. Shastry, K. Wohlfeld, and K. Wu.  We acknowledge support from the U. S. Department of Energy, Office of Basic Energy Science, Division of Materials Science and Engineering, under Contract No. DE-AC02-76SF00515 and the Computational Materials and Chemical Sciences Network (CMCSN) under contract no. DE-SC0007091. Y.W. and C.J.J. were also supported by the Stanford Graduate Fellows in Science and Engineering. A portion of the computational work was performed using the resources of the National Energy Research Scientific Computing Center supported by the US Department of Energy, Office of Science, under contract no. DE-AC02-05CH11231.
\end{acknowledgments}

\bibliographystyle{apsrev4-1}

\end{document}